\documentclass[12pt]{article}
\usepackage{epsfig} 
\usepackage{amssymb}
 
 \def\cf{\hbox{\it cf.}{}}

\def\bra#1{\left\langle #1\right|}
\def\ket#1{\left| #1\right\rangle}

\def\pr#1{#1^\prime}

\def\beq{\begin{equation}}
\def\eeq{\end{equation}}

\def\beqn{\begin{eqnarray}}
\def\eeqn{\end{eqnarray}}
\relax

\catcode`\@=11

\@addtoreset{equation}{section}
\def\theequation{\thesection.\arabic{equation}}

\def\@normalsize{\@setsize\normalsize{15pt}\xiipt\@xiipt
\abovedisplayskip 14pt plus3pt minus3pt%
\belowdisplayskip \abovedisplayskip
\abovedisplayshortskip \z@ plus3pt%
\belowdisplayshortskip 7pt plus3.5pt minus0pt}

\def\small{\@setsize\small{13.6pt}\xipt\@xipt
\abovedisplayskip 13pt plus3pt minus3pt%
\belowdisplayskip \abovedisplayskip
\abovedisplayshortskip \z@ plus3pt%
\belowdisplayshortskip 7pt plus3.5pt minus0pt
\def\@listi{\parsep 4.5pt plus 2pt minus 1pt
     \itemsep \parsep
     \topsep 9pt plus 3pt minus 3pt}}
\@twosidetrue

\catcode`@=12

\newlength{\capwidth}
\catcode`\@=11

\def\section{\@startsection{section}{1}{\z@}{3.5ex plus 1ex minus
   .2ex}{2.3ex plus .2ex}{\large\bf}}

\def\thesection{\arabic{section}}

\def\appendix{\setcounter{section}{0}
 \def\thesection{Appendix \Alph{section}:}
 \def\theequation{\Alph{section}.\arabic{equation}}
  }

\catcode`\@=12

\def \QQ {Q \overline Q}

\def    \be             {\begin{equation}}
\def    \ee             {\end{equation}}
\def    \ba             {\begin{eqnarray}}
\def    \ea             {\end{eqnarray}}
\def    \nn             {\nonumber}
\def    \=              {\;=\;}
\def    \frac           #1#2{{#1 \over #2}}

\def    \bra#1          {\mbox{$\langle #1 |$}}
\def    \ket#1          {\mbox{$| #1 \rangle$}}

\def    \o              {\ifmmode {\cal{O}} \else ${\cal{O}}$ \fi}
\def    \q              {\ifmmode {\cal{Q}} \else ${\cal{Q}}$ \fi}
\def    \oo              {\ifmmode \overline{\cal{O}} \else 
                            $\overline{\cal{O}}$ \fi}
\def    \oneSzero       {\ifmmode {^1S_0} \else $^1S_0$ \fi}
\def    \threeSone      {\ifmmode {^3S_1} \else $^3S_1$ \fi}
\def    \onePone        {\ifmmode {^1P_1} \else $^1P_1$ \fi}
\def    \threePJ        {\ifmmode {^3P_J} \else $^3P_J$ \fi}
\def    \threePzero     {\ifmmode {^3P_0} \else $^3P_0$ \fi}
\def    \threePone      {\ifmmode {^3P_1} \else $^3P_1$ \fi}
\def    \threePtwo      {\ifmmode {^3P_2} \else $^3P_2$ \fi}

\newcommand     \MSB            {\ifmmode {\overline{\rm MS}} \else 
                                 $\overline{\rm MS}$  \fi}
\def    \muf            {\mbox{$\mu_{\rm F}$}}

\def    \mur            {{\mbox{$\mu$}}}

\def    \mul            {{\mu_\Lambda}}

\def    \as             {\mbox{$\alpha_s$}}

\def    \eps            {\ifmmode \epsilon \else $\epsilon$ \fi}
\def    \epsbar         {\ifmmode \bar\epsilon \else $\bar\epsilon$ \fi}
\def    \epsir          {\ifmmode \epsilon_{\rm IR} \else $\epsilon_{\rm IR}$ \fi}
\def    \epsuv          {\ifmmode \epsilon_{\rm UV} \else $\epsilon_{\rm UV}$ \fi}

\def    \Atot           {{\rm A_{tot}}}

\def    \Cggg           {\frac{1}{2M} \frac{\Phi_{(2)}}{3!}\frac{N}{K}}
\def    \Cggg2          {\frac{1}{2M} \frac{\Phi_{(2)}}{2!}\frac{N}{K}}

\def    \pgg#1          {P_{gg}(#1)}
\def    \cpgq#1         {{\cal{P}}_{gq}(#1)}
\def    \cpgg#1         {{\cal{P}}_{gg}(#1)}
\def    \cpqq#1         {{\cal{P}}_{qq}(#1)}

\def    \sp#1#2         {#1#2}       
\def    \eik#1          { \frac{#1 \epsilon_c}{#1 k} }

\def \QQ {Q \overline Q}
\def \qq {\mbox{$q \overline q$}}

\def \s0 {\mbox{$\sigma_{0} $}}

\def    \eps            {\ifmmode \epsilon \else $\epsilon$ \fi}
\def \cf {{\ifmmode C_F{}\else $C_F$ \fi}}
\def \ca {{\ifmmode C_A {}\else $C_A$ \fi}}
  
\def \tf {{\ifmmode T_F {}\else $T_F$ \fi}}
\def \nf {{\ifmmode n_f {}\else $n_f$ \fi}}
\def \nc {{\ifmmode N_c {}\else $N_c$ \fi}}
\def \da {{\ifmmode D_A {}\else $D_A$ \fi}}
\def \Bf {{\ifmmode B_F {}\else $B_F$ \fi}}
\def \df {{\ifmmode D_F {}\else $D_F$ \fi}}

\def \feps#1 {f_{\epsilon}(#1)}

\def\szh    {\mbox{$\sigma_0^{H}$}}

\def\psih {\mbox{$^3S_1^{[8]}$}}

\def\b0{\mbox{$b_0$}}
\def\dsh  {\mbox{$\sigma^{H} $}}

\def\qf     {\mbox{$\mu^2_{\rm F}$}}

\def\eulergamma{\gamma_E}

\begin{document}
\begin{titlepage}
\nopagebreak
{\flushright{
        \begin{minipage}{4cm}
        CERN-TH/99-312  \hfill \\
        hep-ph/9910412\hfill \\
        \end{minipage}        }

}
\vfill
\begin{center}
{\LARGE 
{ \bf \sc
Soft-Gluon Resummation in \\Heavy Quarkonium Physics}}
\vskip .5cm
{\large Matteo Cacciari
}
\\
\vskip .1cm
{CERN, TH Division, Geneva, Switzerland} \\
\end{center}
\nopagebreak
\vfill
\begin{abstract}
Soft-gluon resummation within the framework of heavy quarkonium hadroproduction is
considered. A few selected cases are studied in detail. A sizeable increase of
the cross sections with respect to the next-to-leading order predictions with
central factorization/renormalization scale choice can generally be observed. 
Improvements in the dependence of the cross sections on the two
scales, especially when they are kept equal, are also found.
\end{abstract}
\vskip 1cm
CERN-TH/99-312 \hfill \\
October 1999 \hfill
\vfill
\end{titlepage}

\section{Introduction}

In the twilight zone between perturbative and non-perturbative QCD, there lives
heavy quarkonium physics. Indeed, while the large mass ($m \gg \Lambda_{QCD}$)
of the charm and  bottom quarks allows for a perturbative evaluation of their
production cross sections, still the physics of their binding into
observable charmonium and bottomonium states unavoidably involves low-scale,
and therefore non-perturbative, phenomena.

An understanding of a systematic way to disentangle short- and
long-distance effects was reached by Bodwin, Braaten and Lepage in
\cite{bbl}. By making use of the Non-Relativistic QCD (NRQCD)~\cite{CL86} 
effective field
theory, they were able to provide a framework in which heavy quarkonium
calculations could be carried out to, in principle, arbitrary high order.
Non-perturbative soft effects are factored into NRQCD matrix elements, and
perturbative calculations can provide their coefficient functions. Heavy
quarkonium ($H$) production cross sections therefore take the form:
\be                                                                  
d\sigma(H + X) = \sum_n d\hat\sigma(\QQ[n] + X',\mul)\langle{\cal
O}^H[n]\rangle^{(\mul)}\, .
\label{eq-fm}
\ee
In this expression $\QQ[n]$ represents a heavy quark--antiquark pair in a given
colour and spin/orbital/total angular momentum state. Notice that the quantum
numbers $n$  might differ from the ones of the observable quarkonium $H$, and at
the same time $\QQ$ might even be in a colour-octet state: soft gluons
``hidden'' into the non-perturbative matrix elements 
$\langle{\cal O}^H[n]\rangle$ take care of binding the pair and, at the same
time, building up the correct quantum numbers. $\mul$ represents the NRQCD
factorization scale, separating short- and long-distance effects.
The relative importance of the various contributions in eq.~(\ref{eq-fm})  can
be estimated by using NRQCD velocity scaling rules \cite{LMNMH92}, which allow
the truncation of the series  at any given order of accuracy.

The coefficient functions $d\hat\sigma(\QQ[n] + X)$ can be evaluated in
perturbative QCD. Next-to-leading order (NLO) calculations for a wide range of
processes in hadron--hadron collisions have been presented in \cite{pcgmm}.
Results for the partonic total cross sections will only depend on the scaling
variable $x = 4m^2/s$, $s$ being the partonic centre-of-mass energy.

It is apparent from the fully analytic results listed in \cite{pcgmm} that,
when $x\to 1$, i.e. when the partonic threshold is approached, large NLO
contributions can develop and hence spoil the convergence of the perturbative
series. This calls for an all-order resummation of such contributions, which
is precisely the goal of this paper. We will work using techniques developed 
in~\cite{kt,sterman, cdt, c32}, and more recently applied to heavy-quark production
processes in~\cite{cmnt,bcmn}.

In the following sections we shall first review the structure of the NLO results
and the resummation formalism, and then present some numerical results.

\section{Next-to-leading Order Results and Resummation}

In ref. \cite{pcgmm} results for the total production cross sections 
$\hat\sigma(ij\to \QQ[n]+k)$ are given up to NLO. The colliding partons $ij$ can
be gluons, light quark and antiquark, or a quark and a gluon. Analogously, the
outgoing parton $k$ will be either a gluon or a light quark, and results have
been given for the $\QQ$ pair to be either an $S$-state (both scalar and vector,
both colour singlet and octet) or a $^3P_J$ state.

For ease of reference let us write down here the cross sections for the
processes that display the large soft-gluon behaviour.

The total cross section of the process $gg\to \q g$, with $\q$ representing 
the $\QQ$ pair in a given state $n={}^1S_0^{[1,8]},{}^3P_0^{[1,8]},
{}^3P_2^{[1,8]}$, reads
\beqn
\dsh[gg\to \q\, g] &=& \szh[gg\to\q] \left(\delta(1-x) 
+ \frac{\as(\mur)}{\pi}                                
 \Big\{ \;  \Atot [\q] \;   \delta(1-x)\right.\nn\\ 
&+&\left[x {\overline P}_{gg}(x)\log\frac{4m^2}{x\qf} + 2
   x(1-x) P_{gg}(x)\left(\frac{\log(1-x)}{1-x}\right)_{+}  \right.\nn\\
&+&\left. \left. \left.\left(\frac{1}{1-x}\right)_{+} f_{gg}[\q](x)
   \right]\right\}\right) \; .
\label{eq-gg}
\eeqn
In this equation $P_{gg}(x)$ and ${\overline P}_{gg}(x)$ are related to the
gluon--gluon Altarelli--Parisi splitting vertex:
\beqn
&&P_{gg}(x) = 2\ca\left[\frac{x}{1-x}+\frac{1-x}{x}+x(1-x)\right]\\
&&{\overline P}_{gg}(x) = 
2\ca\left[\frac{x}{(1-x)_{+}}+\frac{1-x}{x}+x(1-x)\right]
\eeqn
where the plus-distribution is defined by\footnote{Notice that in 
ref.~\cite{pcgmm} the results are instead written in terms of 
$\rho$-distributions defined by
\ba                                                                       
   \int_{\rho}^{1} \, dx \, \left[d(x)\right]_{\rho} t(x) =
   \int_{\rho}^{1} \, dx \; d(x) \; \left[t(x)-t(1)\right] \nn \; ,
\ea  
with $\rho = 4m^2/S$ and $S$ being the hadronic centre-of-mass energy.
As a consequence, the constants $\Atot [\q]$ to be inserted in the NLO cross
sections given here differ from the ones published in~\cite{pcgmm}, missing
now all the $\beta$-dependent terms ($\beta\equiv\sqrt{1-\rho}$).}
\be                                                                      
   \int_{0}^{1} \, dx \, \left[d(x)\right]_{+} t(x) =
   \int_{0}^{1} \, dx \; d(x) \; \left[t(x)-t(1)\right] \; ;
\ee      
$\szh[gg\to\q]$ represents the Born cross section for the production of the
quarkonium $H$ via the intermediate state $\q$, i.e. according to
eq.~(\ref{eq-fm}),
\be
\szh[gg\to\q] = \hat\sigma_0[gg\to\q] \langle{\cal O}^H(\q)\rangle\; .
\ee
The constants $\Atot [\q]$ and the functions $f_{gg}[\q](x)$ depend, as
indicated, on the particular $\QQ[n]$ state produced. They can be taken,
together with $\hat\sigma_0[gg\to\q]$, from
the results published in \cite{pcgmm}. Finally, $\mur$ and $\muf$ are
respectively the renormalization and factorization scales. It is worth noticing
that the NRQCD factorization scale $\mul$ does not explicitly appear, at this
order, in the cross section for the production of the $\QQ$ pair. It is
therefore not indicated in the NRQCD matrix element either.

The presence of very large corrections near $x=1$ is clearly visible in 
eq.~(\ref{eq-gg}), in the form of $\left( \log(1-x)/(1-x)\right)_+$ and
$\left(1/(1-x)\right)_+$ distributions. Their form can be even more clearly
appreciated by going to Mellin moments space. The Mellin transform is defined by
\be
f_N \equiv \int_0^1 dx\; x^{N-1} f(x)
\ee
and, in the large-$N$ limit (corresponding to the $x\to 1$ one), 
eq.~(\ref{eq-gg}) becomes
\beqn
\sigma^H_N[gg\to \q^{[c]} g] &=& \sigma^H_{0,N}[gg\to \q^{[c]}] 
\left(1 +  \frac{\as(\mur)}{\pi}                                
\Big\{ \;  \Atot [\q^{[c]}] \;\right.\nn\\ 
&+&2\ca \log^2 N + 2\ca \log N \left(2\eulergamma -
\log{{4m^2}\over{\muf^2}}\right) \nn\\
&+& 2\ca\left(\eulergamma^2 + \frac{\pi^2}{6} - 
\eulergamma \log{{4m^2}\over{\muf^2}}\right) \nn\\
&+&\ca\left(\eulergamma + \log N\right)\delta_{c8} \,\,
+\,\, {\cal O}(1/N)\Big\}\Big) \; ,
\label{eq-gg-mellin}
\eeqn
where $\eulergamma = 0.5772...$ is the Euler constant 
and the superscript $^{[c]}$
refers to the colour state (singlet or octet) of the $\QQ$ pair $\q$.
In this form one can easily see the leading double log, due to soft {\sl and}
collinear radiation from the initial light partons, and the subleading 
single logs. The $\delta_{c8}$ in the
last line indicates that this term is only present when $\q$ is in a
colour-octet state, since it is due to soft radiation from a coloured final state.
It is worth noting that, aside for the constant terms in $\Atot [\q]$ and the
colour of the final state, this structure does not depend on what exactly $\q$
is.

Another process displaying large threshold corrections is $q\bar q \to
\psih g$. The total cross section reads
\beqn
\dsh[\qq\to \psih \, g] &=& \szh[\qq\to \psih]
\left(\delta(1-x)+\frac{\as}{\pi}
\left\{\Atot[\psih]
 \delta(1-x)\right.\right.\nn\\
&+&\left[x {\overline P}_{qq}(x)\log\frac{4m^2}{x\qf} +\cf
   x(1-x) + 2 x(1-x) P_{qq}(x)\left(\frac{\log(1-x)}{1-x}\right)_{+}\right]
   \nn\\                      
&+&\left.\left.\left(\frac{1}{1-x}\right)_{+}
   f_{q\bar q}[\psih](x)\right\}\right) \; ,
\eeqn
where
\be
P_{qq}(x) = \cf\left(\frac{1+x^2}{1-x}\right) \qquad {\rm and} \qquad
{\overline P}_{qq}(x) = \cf\frac{1+x^2}{(1-x)_{+}} \; .
\ee
and, as before, $\szh[\qq\to \psih]$, $\Atot [\psih]$ and 
$f_{q\bar q}[\psih](x)$ can be found in~\cite{pcgmm}.
This becomes, in Mellin moments space,
\beqn
\sigma^H_N[q\bar q\to \psih g] &=& \sigma^H_{0,N}[q\bar q\to \psih] 
\left(1 +  \frac{\as(\mur)}{\pi}                                
\Big\{ \;  \Atot [\psih] \;\right.\nn\\ 
&+&2\cf \log^2 N + 2\cf \log N \left(2\eulergamma -
\log{{4m^2}\over{\muf^2}}\right) \nn\\
&+& 2\cf\left(\eulergamma^2 + \frac{\pi^2}{6} - 
\eulergamma \log{{4m^2}\over{\muf^2}}\right) \nn\\
&+&\ca\left(\eulergamma + \log N\right) \,\,+\,\, {\cal O}(1/N)\Big\}\Big) \; .
\label{eq-qq-mellin}
\eeqn
Comparing this equation to eq.~(\ref{eq-gg-mellin}) we can see that 
the structure is 
identical, with the replacement $\ca\to\cf$ for radiation coming from 
initial-state 
quarks rather than gluons. This cross section can also be seen to be free
of an explicit $\mul$ dependence.

Resummation techniques for these large threshold logarithms have been developed
in \cite{kt,sterman,cdt,c32} and more recently applied to heavy quark 
production~\cite{cmnt,bcmn}. Soft gluon resummation for processes also involving 
non-perturbative matrix
elements, namely radiative and semileptonic $B$ decays within Heavy Quark
Effective Theory, have also been considered in \cite{ks,roth}.

Upon inspection, one can see that the soft limits of
the amplitudes projected onto some specific $\QQ[n]$ state 
are identical, up to this order, to the ones for open heavy quark 
production. Such soft limits, derived in \cite{pcgmm}, read, for 
the gluon--gluon channel and for colour-singlet and -octet production
respectively,
\ba
&&\sum_{\rm col,pol}\left|\overline{A^{[1]}_{\rm soft}}\right|^2 = 
4\pi\as C_A {{2ab}\over{(ak)(bk)}} 
\sum_{\rm col,pol}\left|\overline{A^{[1]}_{\rm Born}}\right|^2\; ,\\
&&\sum_{\rm col,pol}\left|\overline{A^{[8]}_{\rm soft}}\right|^2 = 
4\pi\as C_A \left[{{2ab}\over{(ak)(bk)}} - {{2ab}\over{(Pk)^2}}\right]  
\sum_{\rm col,pol}\left|\overline{A^{[8]}_{\rm Born}}\right|^2\; ,
\ea
where $a, b, P, k$ are, respectively, the momenta of the incoming gluons, the
outgoing heavy $\QQ$ pair, and the emitted soft gluon.
A similar expression holds for $\psih$ production in $\qq$ collisions, 
where we find
\be
\sum_{\rm col,pol}\left|\overline{A^{[8]}_{\rm soft}}\right|^2 = 
4\pi\as \left[C_F {{2q\bar q}\over{(qk)(\bar qk)}} - 
C_A {{2q\bar q}\over{(Pk)^2}}\right]  
\sum_{\rm col,pol}\left|\overline{A^{[8]}_{\rm Born}}\right|^2\;,
\ee
$q$ and $\bar q$ being the momenta of the two incoming light quarks.

Making use of these soft matrix elements one can easily reproduce 
the large-$N$ limits given above. Moreover, one can then achieve 
next-to-leading log (NLL) resummation for heavy quarkonium production processes 
by means of the following formula:
\be
\hat\sigma_N^{res}[ij\to \q^{[c]} g] = \hat\sigma_{0,N}[ij\to \q^{[c]}] \,
                   \left(1 + \frac{\as(\mur)}{\pi}  C_{ij}[\q^{[c]}] \right)\,
                   \Delta_{ij,c,\,N+1}\left(\as(\mur),\mur,\muf\right)
                   \label{resummed}
\ee
the $C_{ij}[\q^{[c]}]$ being the constant, $N$-independent terms readable from
eqs.~(\ref{eq-gg-mellin}) and (\ref{eq-qq-mellin}):
\be
C_{ij}[\q^{[c]}] = \Atot [\q^{[c]}] 
+ 2\left(\begin{array}{c}\ca \\ \cf \end{array}\right)
\left(\eulergamma^2 + \frac{\pi^2}{6} - 
\eulergamma \log{{4m^2}\over{\muf^2}}\right) 
+\ca\, \eulergamma \,\, \delta_{c8} \; ,
\ee
where the coefficient of the second term is either $\ca$ or $\cf$ for $ij = gg$
or $\qq$ respectively, and the last term is only present when a
colour octet state is produced.

The resummation function 
$\Delta_{ij,c,\,N}$ up to NLL accuracy reads
\ba                              
\Delta_{ij,c, \,N} \left( 
\as(\mur), \mur,\muf \right) &=& \exp \left\{ \ln N
\; g^{(1)}_{ij} ( b_0 \, \as(\mur) \ln N) + 
g^{(2)}_{ij,c} (b_0 \, \as(\mur) \ln N, \mur,\muf) \right\} \;\;.
\label{dnexp} 
\ea
The functions $g^{(1)}$ and $g^{(2)}$ resum the LL and NLL terms, respectively.
Their explicit form is \cite{cmnt,bcmn}
\be
g^{(1)}_{q{\bar q}}(\lambda) = \frac{C_F}{\pi b_0 \lambda}
\left[ 2\lambda + (1-2\lambda) \ln (1-2\lambda) \right] \;, \quad 
g^{(1)}_{gg}(\lambda) = \frac{C_A}{C_F} \, g^{(1)}_{q{\bar q}}(\lambda)
\;, \label{g1fun}                         
\ee
and
\beqn
g^{(2)}_{q{\bar q},{ 1}}(\lambda,\mur,\muf) &=&
- \gamma_E \frac{2C_F}{\pi b_0} \ln (1-2\lambda)
+\frac{C_F  b_1}{\pi b_0^3}
\left[ 2\lambda + \ln (1-2\lambda) + \frac{1}{2} \ln^2 (1-2\lambda) \right]
\nn \\
&-& \frac{C_F K}{2\pi^2 b_0^2} \left[2\lambda + \ln (1-2\lambda) \right] 
- \frac{C_F}{\pi b_0} \ln (1-2\lambda) \ln \frac{\muf^2}{4m^2} \nn \\
&-& \frac{C_F}{\pi b_0} \left[\ln (1-2\lambda) + 2\lambda\right] 
\ln \frac{\mur^2}{\muf^2}
 \;,  \nn \\
g^{(2)}_{gg,{ 1}}(\lambda,\mur,\muf) &=& \frac{C_A}{C_F} \,
g^{(2)}_{q{\bar q},{ 1}}(\lambda,\mur,\muf) \;, \label{g2fun} \nn \\
g^{(2)}_{ij,{ 8}}(\lambda,\mur,\muf) &=&
g^{(2)}_{ij,{ 1}}(\lambda,\mur,\muf) 
- \frac{C_A}{2\pi b_0} \ln (1-2\lambda) \;, 
\eeqn
where $b_0, \,b_1$ are the first two coefficients of the QCD
$\beta$-function\footnote{It is worth mentioning the factor 
of $2\pi$ difference between this definition of $b_0$ and the one 
employed in ref.~\cite{pcgmm}.}:
\beq
\label{betacoef}
b_0 = \frac{11C_A - 4T_F n_f}{12 \pi} \;, \quad 
b_1 = \frac{17C_A^2 - 10 C_A T_F n_f - 6 C_F T_F n_f}{24 \pi^2} \;\;,
\eeq
and $K$ is given by
\be
K= \left(\frac{67}{18} - \frac{\pi^2}{6} \right) C_A - \frac{10}{9} T_F n_f \;.
\ee 
It is easy to see that expanding these formulas up to order $\as$, and using
eq.~(\ref{resummed}), one recovers the fixed-order results given in 
eqs.~(\ref{eq-gg-mellin}) and (\ref{eq-qq-mellin}).

Finally, observable hadron-level cross sections will be obtained by convoluting
eq.~(\ref{resummed}) with hadronic parton distribution functions and multiplying
by the proper NRQCD non-perturbative matrix element. We shall therefore have
\be
\sigma^{H,res}_N[\q] = F_{i,N+1}(\muf)\,F_{j,N+1}(\muf)\, 
\hat\sigma_N^{res}[ij\to\q\, g] \, \langle{\cal O}^H(\q)\rangle
\label{res-mellin}
\ee
and an improved hadronic cross section, including the full NLO result plus NLL
resummation, will be obtained as
\beqn
\sigma^{H,\,\mathrm{NLO+NLL}}[\q] = \sigma^{H,res}[\q] - 
(\sigma^{H,res}[\q])_{\alpha_s^3} + \sigma^{H,\,\mathrm{NLO}}[\q]\; ,
\label{matched}
\eeqn
where $\sigma^{H,\,\mathrm{NLO}}[\q]$ is the full NLO result as calculated in
\cite{pcgmm} and $\left(\sigma^{res}[ij\to\q\, g]\right)_{\alpha_s^3}$
is the order-$\alpha_s^3$ truncation of the resummed result, here subtracted to 
avoid double-counting.

\section{Numerical Results}

Phenomenological cross sections are obtained via inverse Mellin transform
of eq.~(\ref{resummed}), to be performed numerically. Many problems are 
to be found at this stage, both of conceptual and technical nature.

For instance, to take care of the presence of the Landau pole  we shall adopt
the so-called Minimal Prescription introduced in \cite{cmnt}: the integration
contour for the inverse Mellin transform passes to the right of all
singularities in the complex $N$-plane, but to the left of the Landau pole
at $N = N_L \equiv \exp(1/2 b_0 \as)$.

Technical problems are instead related to the difficulty of
performing the required integrals, due to the resummation function $\Delta(x)$ 
being strongly oscillating 
in $x$-space close to $x=1$. The are taken care of by a subtraction procedure
similar to the one described in Appendix B of ref.~\cite{cmnt}.

In the following we shall present some plots showing the effect of the 
soft-gluon resummation described in the previous section. Hadron level results
are obtained with the CTEQ3M parton distribution function (PDF) set, unless
otherwise stated. NRQCD matrix elements for $\QQ$ level final states (i.e. not
real physical quarkonium states) are fixed, for $S$-states, at the value number
of polarization states $\times$ number of colours of the $\QQ$ pair. If
replaced with the real values, in units of GeV$^3$ for $S$-states and GeV$^5$
for $P$-states, the cross sections would be in nanobarns.

We shall first study the renormalization/factorization scale dependence of a
fictitious heavy quarkonium made up of 50 GeV quarks. Charmonium and
bottomonium are very close to the non-perturbative region, and the large value
of the strong coupling at such low scales makes the results less readily
readable.

\begin{figure}[!tb]
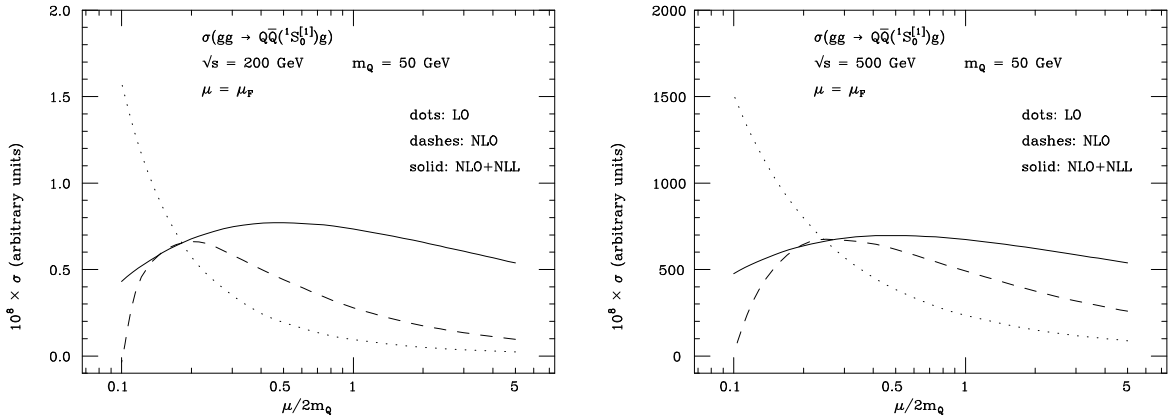

\begin{center}
\begin{minipage}{8cm}
\flushright
\epsfig{file=m50-15-200.ps,height=5.5cm}
\end{minipage}
\begin{minipage}{8cm}
\flushright
\epsfig{file=m50-15-500.ps,height=5.5cm}
\end{minipage}
\parbox{\capwidth}{
\caption{\label{fig:m50} \protect\small
The cross section for production, in $pp$ collisions via the gluon-gluon
channel, 
of a fictitious ${}^1S_0^{[1]}$ quarkonium state made up of 50 GeV quarks, 
at 200 and 500 GeV 
centre-of-mass energy.
}}
\end{center}
\end{figure}

In fig.~\ref{fig:m50} resummation effects are shown at 200 and 500 GeV
centre-of-mass energies, as a function of the renormalization/factorization
scale $\mu$. Production of a ${}^1S_0^{[1]}$ state in proton--proton collisions
via the gluon--gluon channel is considered here.
While the leading order result displays the usual monotonical
dependence, and the NLO one presents a maximum at pretty small values of $\mu$,
the resummed result can be seen to be markedly less dependent on the arbitrary
scale. One can also see resummation effects to be larger closer to the
threshold, at $\sqrt{S} = 200$~GeV, than at 500 GeV, as expected. In 
both cases a sizeable
increase with respect to the NLO prediction with $\mu = 2m_Q$ can be
observed.

Results for other states produced via the gluon--gluon channel, i.e.
$\threePzero$ and $\threePtwo$, both in the colour singlet and octet states,
and ${}^1S_0^{[8]}$, appear qualitatively similar. These plots 
 suggest that a small value of the renormalization/factorization scale
should be chosen in the NLO calculation in order to obtain a more realistic
prediction for the cross section. We can also speculate that the value
$\mu=m_Q$, rather than $2m_Q$, should be chosen as the central one for the
renormalization and factorization scale, despite the $\log(4m^2/\mu^2)$ terms
appearing in the NLO calculation. A variation in the range $[m_Q/2,2m_Q]$ would
then still span a large fraction of the NLO band and give a pretty small 
uncertainty with the NLO+NLL result.

\begin{figure}[!tb]
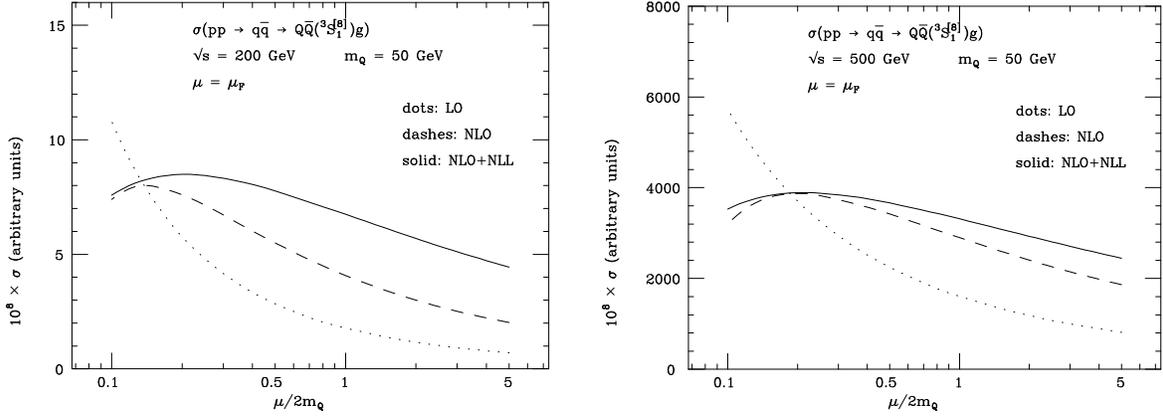

\begin{center}
\begin{minipage}{8cm}
\flushright
\epsfig{file=m50-66-200.ps,height=5.5cm}
\end{minipage}
\begin{minipage}{8cm}
\flushright
\epsfig{file=m50-66-500.ps,height=5.5cm}
\end{minipage}
\parbox{\capwidth}{
\caption{\label{fig:m50-qq} \protect\small
The cross section for production, in $pp$ collisions via the $q\bar q$
channel, of a ${}^3S_1^{[8]}$ state with 50 GeV quarks.
}}
\end{center}
\end{figure}

Qualitatively similar results are found in the $q\bar q$ channel for production
of a ${}^3S_1^{[8]}$ state. Figure~\ref{fig:m50-qq} shows the dependence of the
cross section on the renormalization/factorization scale at $\sqrt{S} = 200$
and 500 GeV. Again, the NLO+NLL result can be seen to be more stable than the
fixed-order NLO one.

\begin{figure}[!tb]
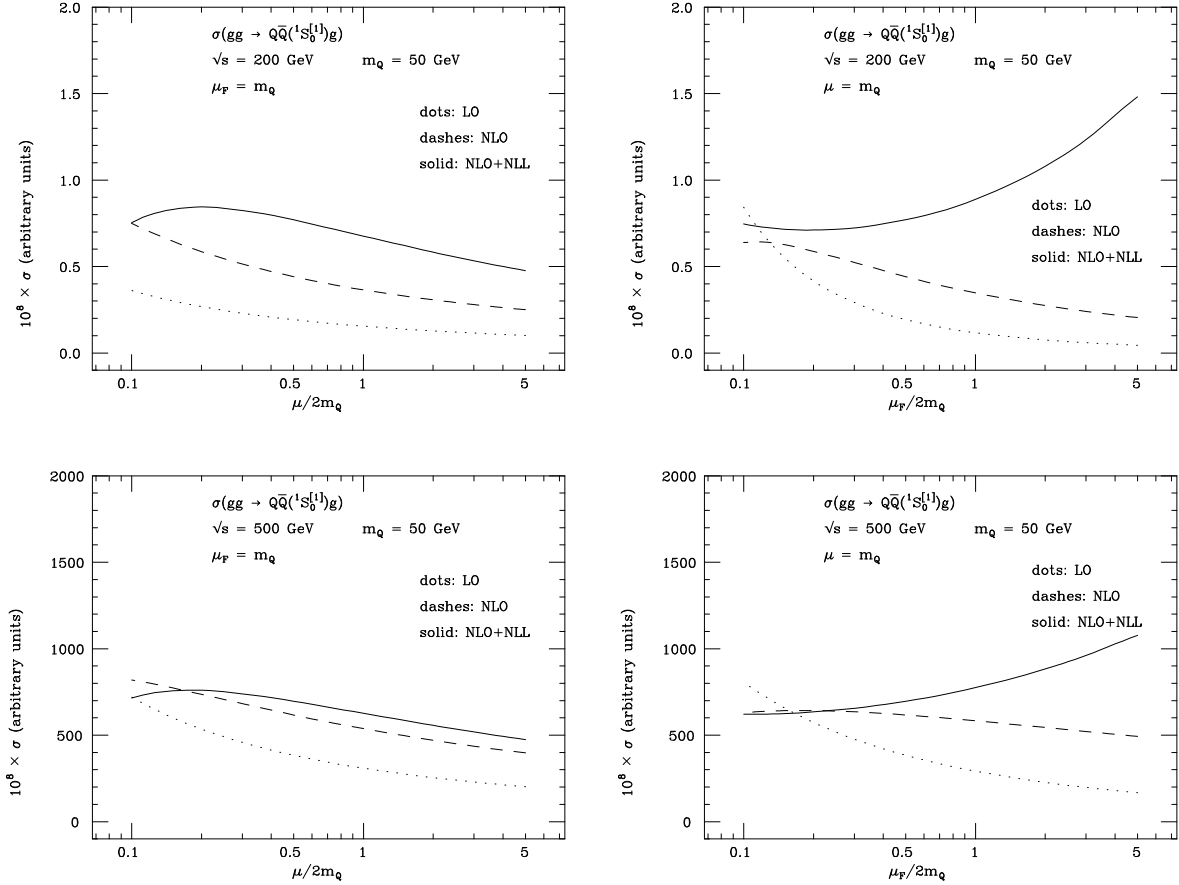

\begin{center}
\begin{minipage}{8cm}
\flushright
\epsfig{file=m50-15-200-varmur.ps,height=5.5cm}
\end{minipage}
\begin{minipage}{8cm}
\flushright
\epsfig{file=m50-15-200-varmuf.ps,height=5.5cm}
\end{minipage}\\[20pt]
\begin{minipage}{8cm}
\flushright
\epsfig{file=m50-15-500-varmur.ps,height=5.5cm}
\end{minipage}
\begin{minipage}{8cm}
\flushright
\epsfig{file=m50-15-500-varmuf.ps,height=5.5cm}
\end{minipage}
\parbox{\capwidth}{
\caption{\label{fig:m50-murmuf} \protect\small
Cross section as in fig.~\protect\ref{fig:m50}, but with independent variation
of the renormalization (left plot) and factorization (right plot) scales. Top
plots are for $\sqrt{S} = 200$ GeV, bottom ones for $\sqrt{S} = 500$ GeV.
}}
\end{center}
\end{figure}

One can, of course, also try to vary the renormalization scale $\mur$ and the
factorization scale $\muf$ independently. According to the observation made
above on the ``best'' scales central value, we now fix one of the scales at
$m_Q$ (rather than $2m_Q$) and we vary the other in the range $[2m_Q/10,
5\times 2m_Q]$. We can see  in  fig.~\ref{fig:m50-murmuf} that the remarkable
independence seen in fig.~\ref{fig:m50} is unfortunately at least partially
lost, especially when going to large factorization scales. Independent scale
variations in the $\qq$ channel give a similar outcome. These results are
qualitatively similar to those reported in \cite{cmnov}, where soft-gluon
resummation in prompt-photon hadroproduction is studied, indicating that they
are not specific to bound states production. They are moreover identical to
what can be obtained by making $m_Q$ even as large as the top quark mass, i.e.
175 GeV. This suggests that, when studying the overall dependence on the
renormalization and factorization scales, one should always take care to vary
them independently. 

We now move to studying real quarkonium states, i.e. made up of charm or bottom
quarks. In these cases, and especially for charmonium, the not-so-large mass of
the heavy quark, and hence the relatively large value of the running coupling at
these scales, will produce less clear-cut results than the ones previously
displayed.

\begin{figure}[!tb]
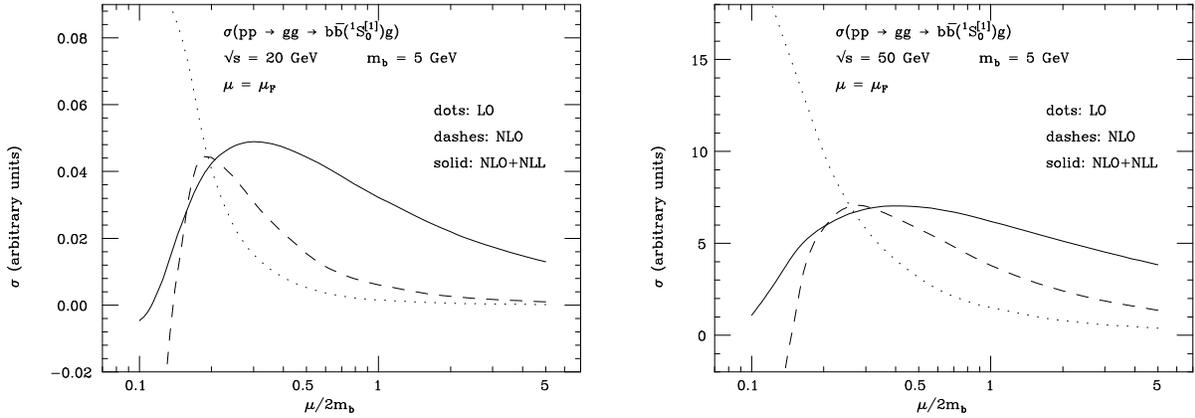

\begin{center}
\begin{minipage}{8cm}
\flushright
\epsfig{file=bottom-15-20.ps,height=5.5cm}
\end{minipage}
\begin{minipage}{8cm}
\flushright
\epsfig{file=bottom-15-50.ps,height=5.5cm}
\end{minipage}
\parbox{\capwidth}{
\caption{\label{fig:bottom} \protect\small
The cross sections for production of the bottomonium state ${}^1S_0^{[1]}$ via
the gluon--gluon channel, at 20 and 50 GeV centre-of-mass energy, as functions 
of the renormalization/factorization scales.
}}
\end{center}
\end{figure}
Figure~\ref{fig:bottom} is a scaled-down version of fig.~\ref{fig:m50},
meaning that both the mass of the heavy quark and the centre-of-mass energy are
scaled down by a factor of 10, to yield $m_b = 5$~GeV and $\sqrt{S} = 20$ 
and 50~GeV.  We can see that, while
the behaviour with respect to scale variations is the same,
even after resumming the soft-gluon logs the cross section does however 
retain a larger dependence than in the 
$m_Q = 50$~GeV case, as expected. As before, the NLO+NLL result predicts larger
cross sections than the NLO one with a central scale choice.

Once again, qualitatively similar results are found for
$\threePzero$ and $\threePtwo$, both in the colour singlet and octet states,
and for ${}^1S_0^{[8]}$.

\begin{figure}[!tb]
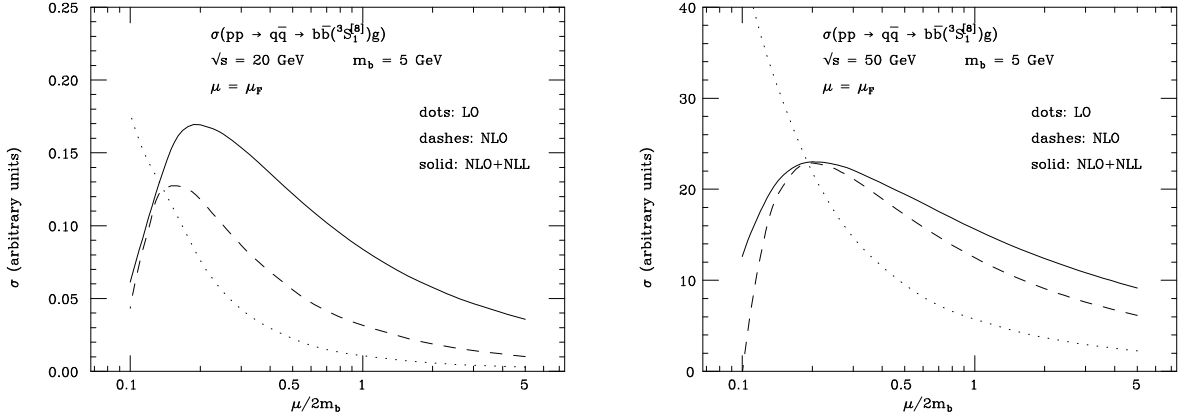

\begin{center}
\begin{minipage}{8cm}
\flushright
\epsfig{file=bottom-66-20.ps,height=5.5cm}
\end{minipage}
\begin{minipage}{8cm}
\flushright
\epsfig{file=bottom-66-50.ps,height=5.5cm}
\end{minipage}
\parbox{\capwidth}{
\caption{\label{fig:bottom-qq} \protect\small
The cross sections for production, in $pp$ collisions via the $q\bar q$
channel, of a ${}^3S_1^{[8]}$ bottomonium state.
}}
\end{center}
\end{figure}
Production of a ${}^3S_1^{[8]}$ bottomonium state in $q\bar q$ interactions can
also be considered. 
Figure~\ref{fig:bottom-qq} shows the dependence of the cross section on the
renormalization/factorization scale at $\sqrt{S} = 20$ and 50 GeV. At 50 GeV
centre-of-mass energy the NLO+NLL
result can be seen to be slightly more stable than the fixed-order NLO one,
confirming the trend observed in $gg$ collisions. Less is instead to be gained
closer to the threshold, at $\sqrt{S} = 20$ GeV.

\begin{figure}[!tb]
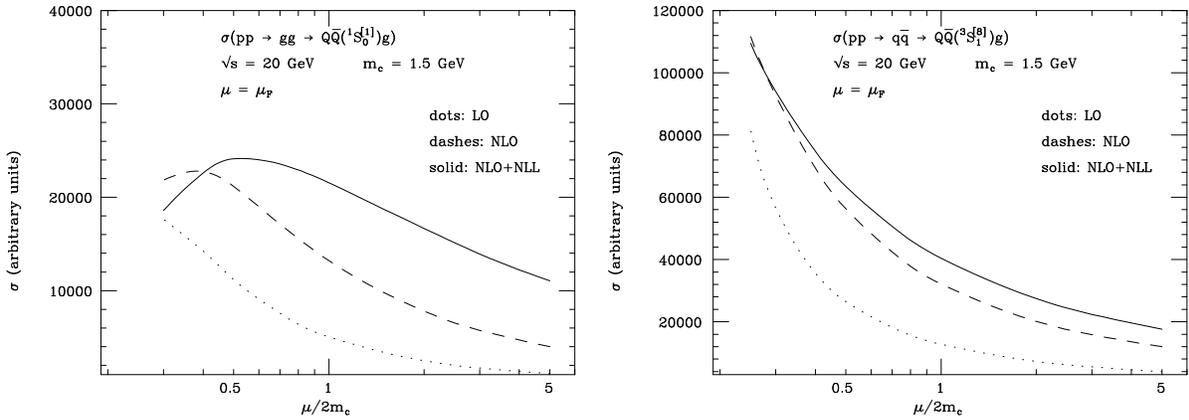

\begin{center}
\begin{minipage}{8cm}
\flushright
\epsfig{file=charm-15-20.ps,height=5.5cm}
\end{minipage}
\begin{minipage}{8cm}
\flushright
\epsfig{file=charm-66-20.ps,height=5.5cm}
\end{minipage}
\parbox{\capwidth}{
\caption{\label{fig:charm} \protect\small
The cross section for production, in $pp$ collisions, of
${}^1S_0^{[1]}$ and ${}^3S_1^{[8]}$ charm bound states. The PDF set is CTEQ4lQ.
}}
\end{center}
\end{figure}
One can finally also try studying charmonium hadroproduction.
Figure~\ref{fig:charm} shows the renormalization/factorization scale dependence
for production of ${}^1S_0^{[1]}$ and ${}^3S_1^{[8]}$ charm bound states in
$pp$ collisions at 20 GeV centre-of-mass energy. The parton distribution
function set CTEQ4lQ (i.e. low $Q^2$, valid down to $Q=0.7$~GeV) is used, since
very small scales are probed. Moreover, the scales are not varied below
$\mu \sim 0.25 \times 2m_c \sim 0.75$~GeV, already
at the borderline of both perturbative QCD and the validity range of the PDF
set.

One can see from the plots that the dependence on the renormalization and
factorization scales is once more lessened when soft gluon resummation is
included. However, the improvement is less than in the bottomonium case, as
expected, the scale involved here being extremely close to the 
non-perturbative region

\section{Open Questions}

This paper only provides a first step towards a full understanding of soft-gluon
effects in heavy quarkonium physics, as a number of questions remain to be
addressed.

The NLO cross sections we have studied here do not present the need for an
explicit subtraction of infrared singularities to be absorbed into the
non-perturbative NRQCD matrix elements. Such singularities are however expected
to appear in higher orders, as they do for instance in the two-loop evaluation
of $J\!/\!\psi$ decay into leptons~\cite{bss}. The resummation function
$\Delta$ will then have to be modified accordingly, so as to reflect the
presence of this new factorization scale $\mu_{\Lambda}$, in the same way it
contains the PDF's  factorization scale $\mu_F$. This will avoid double
counting of contributions  already included somewhere else. It is also
conceivable that the interplay of this subtraction with the one of Coulomb
terms, which  are also absorbed, already at the one-loop level, into the NRQCD
matrix elements, will have to be carefully considered.

In ref. \cite{pcgmm}, where the fixed order NLO cross sections were evaluated,
the need for an explicit subtraction of infrared singularities, and hence the
appearance of the NRQCD factorization scale $\mu_\Lambda$, is however in
one case already met at order $\alpha_s^3$. This happens when studying
the production of a $P$-wave state via the $\qq$
channel. When the emitted gluon becomes soft, the ensuing singularity can only
be cancelled by adding the corresponding $\qq \to S$-state process, in full
accordance with the general NRQCD factorization formula~(\ref{eq-fm}) (see
also section 6 of ref.~\cite{pcgmm} for more details). In this
case we find indeed that the soft limit of the $\qq \to P$-state$\,+g$ process
amplitude does not factorize onto $\qq \to P$-state and no gluons, this cross
section being actually zero, but rather onto the matrix element squared for 
$\qq \to S$-state. This implies a different
framework for achieving soft-gluon resummation, which will conceivably be
closely interlaced with the structure of the NRQCD matrix elements.

\section{Conclusions}

In this paper we have studied the effect of resumming soft gluons in some
selected heavy quarkonium hadroproduction processes. The inclusion of
resummation effects leads in many cases  to sizeably larger cross sections than
those given by the fixed order  NLO result with central
renormalization/factorization scales. Moreover, improvements in the dependence
of the cross sections on the renormalization and factorization scales can be
observed, especially when they are kept equal. We have also pointed out that
less marked improvements appear when we vary them independently, leading to an
uncertainty band a little larger than what would be otherwise obtained.

A full systematic analysis of soft-gluon resummation in heavy quarkonium
processes, with phenomenologically relevant results, is in progress, as is the
one of the open questions briefly described in the previous section.

\vspace{1cm}
\noindent
{\bf Acknowledgements}\newline\noindent
I am most grateful to Stefano Catani for the many conversations we had on 
the subject of soft-gluon resummation. I also thank him and Fabio Maltoni 
for reading the manuscript and providing useful comments.

\end{document}